# Active Contrast-Dissipation All-Optical Encryption

DONGYANG DU[1,#], ZIXUAN FU[1,#], FEI DAI[2,*] AND HANG CHEN[2,*]

[1]School of Instrumentation Science and Opto-electronics Engineering, Beihang University, Beijing, 100191, China

[2]School of Electronics and Information Engineering, Beihang University, Beijing, 100191, China

[#]These authors contributed equally to this work.

*Corresponding author: daphige@buaa.edu.cn (F. Dai); chen_hang@buaa.edu.cn (H. Chen)

**Abstract.** Optical encryption is meant to place secrecy in physical structure, yet adversaries can readily construct imperfect physical keys to probe the encryption system via trial-and-error iterations. Such systems inevitably leak target plaintext information and preserve invertible residual structural features, rendering the entire encryption scheme vulnerable to adversarial inversion attacks. Here, we report an active contrast-dissipation all-optical encryption architecture, which partitions wavefront modulation into $M$ wavelength-private encryption layers and $N$ shared decoding layers. Security rests on multi-dimensional physical keys that must match simultaneously, including illumination wavelength, precise spatial alignment and layer stacking sequence. With every key matched and authorized readout is preserved: the composite plaintext is reconstructed at a full-detector Pearson correlation coefficient (PCC) of 0.997. We further impose active contrast dissipation as a core security constraint. Under physical perturbation it suppresses the foreground-background intensity contrast, so that the slightest key mismatch collapses the target features and dissipates the plaintext into randomized, diffuse speckle within chaotic background noise, leaving no residual semantic leakage to invert. Under nanoscale wavelength detuning ($|\Delta\lambda| \geq 2.5$ nm), a 1° in-plane rotation, a single-pixel lateral displacement or layer-stacking tampering, the plaintext vanishes and the PCC approaches zero, whereas conventional diffractive encryptors retain identifiable contours. This work provide a feasible and reliable solution for high-security information encryption, with great application potential in communication security, information anti-counterfeiting, and confidential data storage.

**Keywords:** All-Optical Encryption, Active Contrast-Dissipation, Photonic Neural Networks.

## 1 Introduction

The rapid growth in data throughput driven by emerging technologies such as 6G communications [1] and large-scale Internet of Things (IoT) deployments [2] has made the secure transmission of confidential data and hardware-level protection core scientific challenges [3]. Traditional electronic cryptography relies primarily on techniques such as symmetric block ciphers (e.g., Advanced Encryption Standard, AES [4]), public-key schemes (e.g., Rivest-Shamir-Adleman, RSA [5]) and so on [6, 7]. However, as computational capabilities increase, emerging quantum algorithms such as Shor's algorithm [8] pose long-term security concerns for traditional electronic cryptography system proposed above. In contrast, optical computing, particularly photonic neural networks (PNNs [9, 10]), utilizing photons as information carriers, offers numerous advantages such as signal processing capabilities based on light propagation, spatial parallelism, and the potential for high energy efficiency. By leveraging multiple optical degrees of freedom (e.g., phase [11], wavelength [12], polarization [13], and so on [14, 15]) to construct high-dimensional key spaces, and by incorporating the physical uniqueness of micro-/nano- structures [16], optical computing architectures provide a highly promising physical implementation scheme for high-security cryptographic applications.

However, deploying PNNs for high-security cryptographic applications still suffer from critical security vulnerabilities and inherent optimization drawbacks that hinder practical secure deployment. Most existing PNN-based optical encryption frameworks adopt passive forward optimization dominated [17] by single-objective fidelity metrics based on mean squared error loss function (MSE) or softmax cross-entropy (SCE) loss function, and lack dedicated active defense mechanisms to resist prevalent unauthorized physical interference [18]. These reconstruction-centric optimization strategies only prioritize the decoding fidelity of authorized outputs, while imposing no explicit constraints on output characteristics under physical perturbations. Such passive optimization paradigms inherently fail to suppress residual semantic leakage: encrypted targets can retain recognizable structural contours under minor physical disturbances (e.g., nano-scale wavelength shifts [19], spatial alignment deviations [20], or physical layer permutations), rendering encrypted patterns susceptible to recovery by adversaries. Although current multi-wavelength optical encryption architectures have improved basic encryption performance, they generally adopt a fully coupled design where all wavelengths share unified and fixed physical modulation layers [18]. This tightly coupled structure cannot support hierarchical access control and differentiated encryption for multi-wavelength composite plaintexts, fails to isolate wavelength-private encryption information from shared decoding channels, and substantially restricts the flexibility and security of optical encryption.

To address the aforementioned security vulnerabilities and optimization limitations of existing PNN-based optical encryption systems, this paper proposes an active contrast-dissipation all-optical encryption framework built upon a specially designed “M+N” decoupled multi-wavelength diffractive deep neural network architecture [9, 12]. Different from conventional fully-connected multi-wavelength encryption designs, the proposed system decomposes the encryption process into $M$ wavelength-private encryption layers and $N$ shared decoding layers, which collaboratively implement all-optical encryption and enable fine-grained hierarchical access control for multi-wavelength composite plaintexts. In this configuration, the detector captures the incoherent superposition of wavelength-resolved intensity distributions, while all wavelength channels maintain inherent physical correlations via shared height-modulated encryption units. Rather than adopting the prevalent passive single-objective forward-fidelity optimization that only guarantees authorized decryption performance, this work proposes a special designed active contrast-dissipation mechanism as the core security constraint. Specifically, the introduced active contrast-dissipation mechanism actively suppresses the foreground-background intensity difference under physical perturbations, which effectively collapses the feature contrast of target patterns and yields randomized, diffuse speckle outputs for unauthorized access attempts. The system constructs multi-dimensional physical keys including illumination wavelengths, precise spatial alignment and stacking sequence of wavelength-private encryption layers and shared decoding layers, and encrypts identity authentication information and multi-wavelength composite patterns as target plaintexts. When facing typical unauthorized physical disturbances, including nano-scale wavelength shifts, spatial alignment deviations, and physical layer permutations, the mismatch of optical propagation conditions will trigger the active contrast-suppression effect, resulting in drastically degraded target contrast and negligible residual semantic recoverability. This work provides an effective physical-layer information rejection paradigm to eliminate covert security risks. The proposed active contrast-dissipation all-optical encryption architecture holds promising application prospects in high-security information storage, secret-sharing-based cooperative image reconstruction, and high-performance computing systems.

## 2 Results

### 2.1 Encryption and Reconstruction of Authorized Information

We proposed a high-security all-optical encryption system based on a "M+N" decoupled multi-wavelength diffractive deep neural network architecture as shown in Figure 1. This architecture physically partitions wavefront modulation into $M$ wavelength-private encryption layers and $N$ shared decoding layers, where each height-modulated encryption unit is trained and modeled using the angular-spectrum method discussed in Section 4.1. In this study, we implement a configuration of $M$=3 and $N$=2.

To enhance system security, we implement a dual-unlock experiment starting with identity (ID) verification as shown in Figure 1a. In the ID verification, three IDs are analogous to three senior managers who hold keys to a bank vault: the vault can only be opened when all three managers are present simultaneously. Each wavelength-private encryption layer named $L_1$, $L_2$ and $L_3$ is assigned to a specific coherent illumination wavelength $\lambda_1$=635 $nm$, $\lambda_2$=532 $nm$ and $\lambda_3$=450 $nm$ respectively. Each wavelength propagates along its own dedicated optical path through one wavelength-private encryption layer and two shared decoding layers named $L_4$ and $L_5$ as shown in Figure 1a, reconstructing the unique ID patterns named $ID_1$, $ID_2$, and $ID_3$ at the detector plane as shown in Figure 1b. Without loss of generality, three different kinds of IDs are employed in this work: snowflake ($ID_1$), cartoon robot ($ID_2$), and petal ($ID_3$). The Pearson correlation coefficient (PCC) quantifies the linear correlation between the reconstructed intensity distribution and the target image, where a PCC value closer to 1 signifies higher reconstruction fidelity. The definition of this metric and the evaluation procedures are elaborated in Section 4.3. In the ID verification, we find that $ID_1$, $ID_2$, and $ID_3$ only respond to their preconfigured wavelengths $\lambda_1$, $\lambda_2$, and $\lambda_3$, yielding full-detector PCC values of 0.976, 0.969, and 0.982, respectively. We further observe that mismatched wavelengths trigger the active contrast-dissipation mechanism described in Section 4.2 and fail to activate the corresponding IDs as shown in Figure 1b. In this case, no ID patterns are visible to the naked eye, and the PCC approaches nearly zero. This further validates the rationality of the wavelength-private encryption layer design.

The target plaintext can only be unlocked only when all three IDs are present simultaneously, which substantially improves the security performance of the proposed system. As shown in Figure 1c, this reconstruction process is realized as follows: under simultaneous illumination of all three wavelength channels, optical fields propagate sequentially through three wavelength-private encryption layers $L_1$, $L_2$ and $L_3$ followed by two shared decoding layers $L_4$ and $L_5$. Reconstruction succeeds only if the multi-dimensional physical keys including illumination wavelengths, precise spatial alignment, and the stacking sequence of the wavelength-private encryption layers and shared decoding layers are correctly configured. In this case, the intensity distribution of each wavelength channel incoherently superimposes at the detector plane to yield the target composite plaintext. In contrast, the target plaintext corresponds to a pattern of an entirely different category from these IDs. Consisting of a random sequence of Arabic numerals as well as uppercase and lowercase English letters, it comprises 16 characters arranged in a $4 \times 4$ array, with the sequential string A8fH2qS7JK6Wp5Xy. As shown in Figure 1d, the reconstructed composite plaintext attains a full-detector PCC of 0.997. Both visual inspection and image-domain results demonstrate the high fidelity of optical reconstruction of our system.

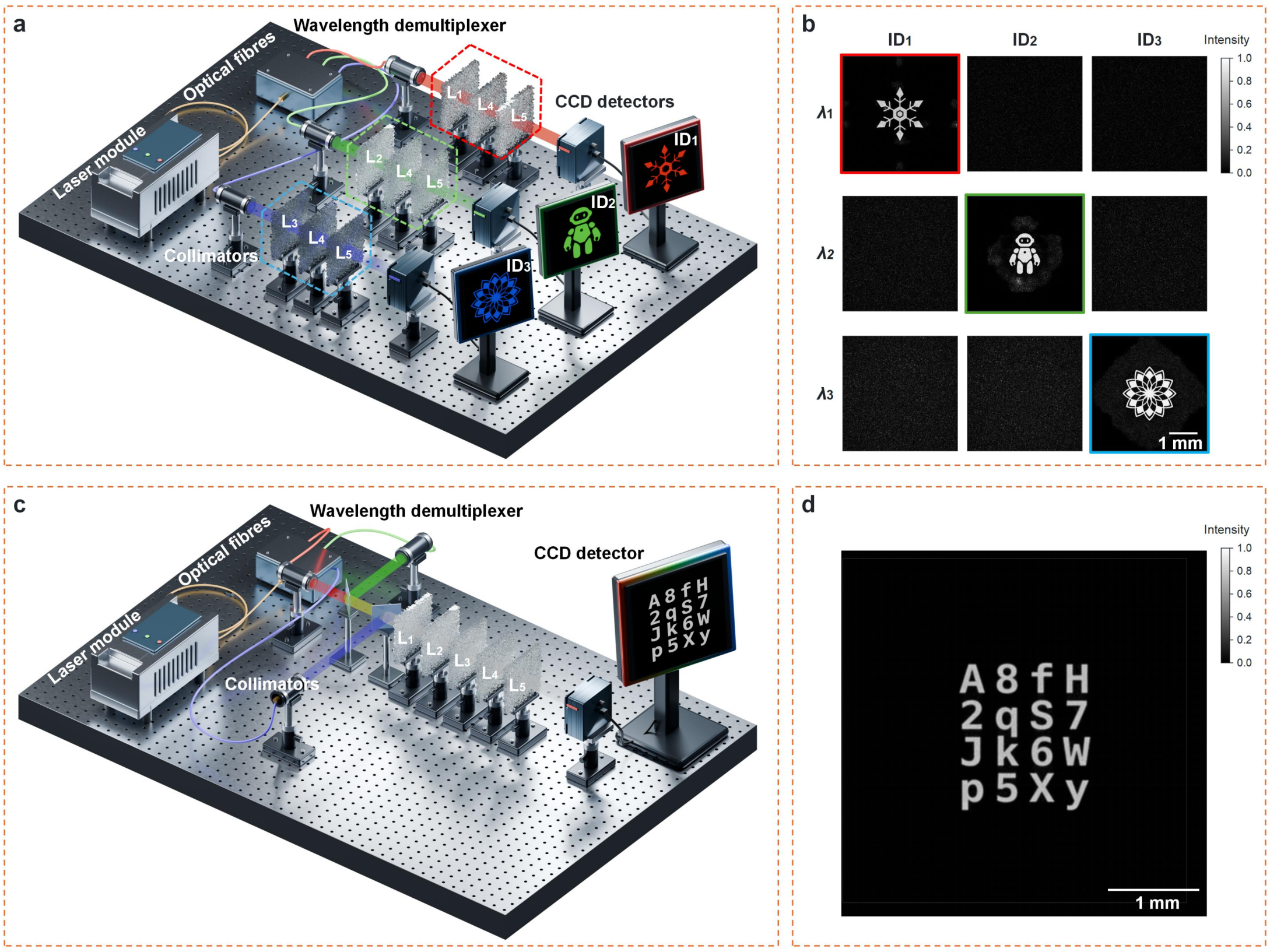

**Figure1. The "M + N" decoupled architecture and encryption/reconstruction of authorized information. (**a) ID verification mode. Three wavelength-stabilised laser modules are combined into optical fibres and separated by a wavelength demultiplexer into three dedicated optical paths, each carrying one wavelength-private encryption layer ($L_1$ at $\lambda_1$ = 635 nm, $L_2$ at $\lambda_2$ = 532 nm, $L_3$ at $\lambda_3$ = 450 nm) and the two shared decoding layers $L_4$ and $L_5$. (b) Detector-plane intensity in the ID verification mode. Rows, illumination wavelength; columns, identity pattern $ID_1$ (snowflake), $ID_2$ (cartoon robot) and $ID_3$ (petal). Each ID is reconstructed only under its pre-assigned illumination wavelength with full-detector PCCs of 0.976, 0.969 and 0.982; illumination at any other wavelength dissipates the plaintext into chaotic background noise and leaves the corresponding ID inactive (PCC$\approx$0). (c) Cooperative composite-plaintext mode, in which the three wavelengths are combined and traverse the full five-layer cascade from $L_1$ to $L_5$. (d) Composite plaintext reconstructed with all multi-dimensional physical keys matched: the 16-character string A8fH2qS7JK6Wp5Xy in a 4×4 array, with a full-detector PCC of 0.997.

## 2.2 Active Contrast-Dissipation Under Unauthorized Physical Keys

In general, valid reconstruction of the target plaintext can only be achieved when all physical keys are perfectly matched. Nevertheless, adversaries commonly attempt to spoof optical encryption systems by randomly probing the physical key space through perturbing the illumination wavelengths, spatial alignment, or layer stacking sequence as shown in Figure 2. Traditional optical encryption architectures are vulnerable to such brute-force probing attacks: mismatched and unauthorized physical key configurations still tend to generate recognizable plaintext patterns or residual plaintext contours discussed in Section 3.2, which greatly facilitates plaintext inversion and cracking. In contrast, benefiting from the proposed active contrast-dissipation mechanism whose details are described in Section 4.2, our system enables rigorous physical-key-dependent reconstruction. Any slight deviation from the authorized physical keys will lead to immediate dissipation of the plaintext information, thereby fundamentally resisting unauthorized probing and spoofing attacks.

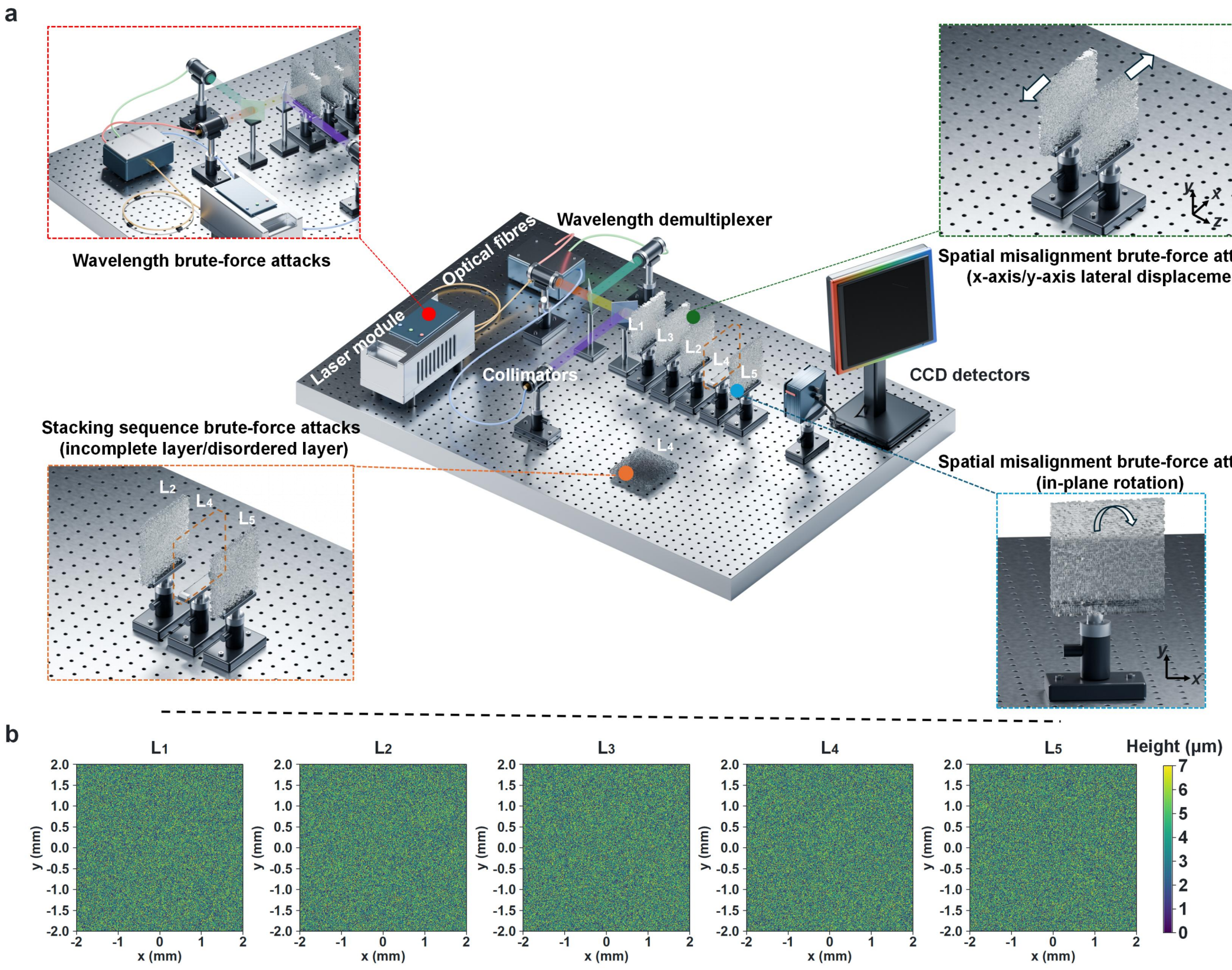

**Figure 2. Authorized physical keys and their typical spoofing attacks.** (a) The multi-dimensional physical keys, which must be matched simultaneously for decryption: the illumination wavelength, the precise spatial alignment of each layer and the stacking sequence of the five layers. Insets, the corresponding brute-force cracking tests: wavelength detuning, in-plane rotation, lateral displacement along the x- and y-directions, and incomplete layer configurations or disordered layer stacking. (b) Height profiles from $L_1$ to $L_5$. The composite plaintext exists only for the exact $L_1$–$L_2$–$L_3$–$L_4$–$L_5$ stacking sequence.

We first perform brute-force cracking tests targeting the illumination wavelengths to validate the effectiveness of the proposed active contrast-dissipation mechanism. The proposed all-optical encryption system adopts three authorized wavelengths, namely $\lambda_1$=635 $nm$, $\lambda_2$=532 $nm$ and $\lambda_3$=450 $nm$. The PCC response curve under wavelength brute-force attacks, plotted in Figure 3a, characterizes the system behavior over the detuning range from −20 nm to 20 nm for all three authorized wavelengths. Further considering that high-stability commercial laser modules generally feature a wavelength tolerance of $|\Delta\lambda| \approx 2.5$ nm [21], any authorized wavelength detuning satisfying $|\Delta\lambda| \geqslant 2.5$ nm is defined as a valid brute-force physical-key attack in this work. The target plaintext vanishes entirely from the detector plane and is invisible to the naked eye, as illustrated in Figure 3b. Correspondingly, the PCC drops sharply from the authorized value of 0.997 at $\Delta\lambda = 0$ to below 0.05 at $\Delta\lambda = \pm 3.0$ nm. This phenomenon originates from the active contrast dissipation effect, which actively dissipates the plaintext information into chaotic background noise without retaining any valid plaintext contours. Such a mechanism thoroughly eliminates residual readable features and effectively prevents adversaries from implementing plaintext inversion and cracking attacks.

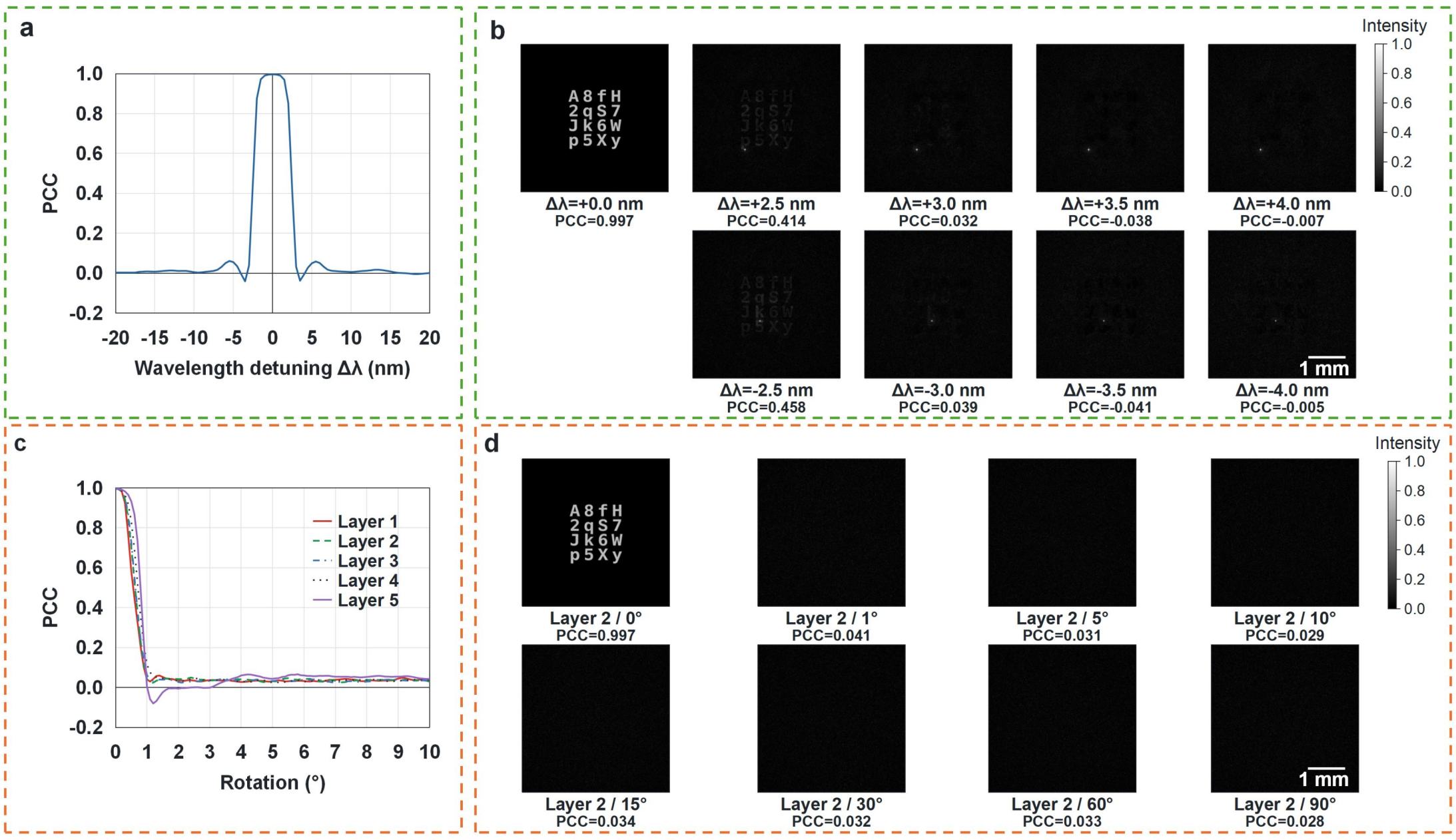

**Figure 3. Active contrast-dissipation under unauthorized physical keys: wavelength detuning and in-plane rotation.** (a) Full-detector PCC of the composite plaintext versus the wavelength detuning $\Delta\lambda$ from −20 nm to +20 nm applied to all three channels. (b) A detuning of $|\Delta\lambda| \geq 2.5$ nm is defined as a valid brute-force physical-key attack; the PCC drops from the authorized value of 0.997 at $\Delta\lambda = 0$ to below 0.05 at $\Delta\lambda = \pm 3.0$ nm and the plaintext is fully dissipated into chaotic background noise for $|\Delta\lambda| \geq 3$ nm. (c) Full-detector PCC versus in-plane rotation from 0 to 10°. (d) Detector-plane intensity for in-plane rotation of $L_2$ by 0°, 1°, 5°, 10°, 15°, 30°, 60° and 90° (PCC = 0.997 at 0° and 0.041, 0.031, 0.029, 0.034, 0.032, 0.033 and 0.028 for the rotated cases).

We further perform brute-force cracking tests targeting the precise spatial alignment to validate the superiority of the proposed active contrast-dissipation mechanism. We primarily focus on two typical spatial misalignment scenarios of the wavelength-private encryption layers and shared decoding layers: in-plane rotation and lateral displacement along the $x$- and $y$-directions. These spatial disturbances are highly feasible for adversarial manipulation and thus serve as accessible brute-force cracking approaches. The PCC response curve under in-plane rotation brute-force attacks, plotted in Figure 3c, characterizes the system behavior over the rotation range from 0° to 10° for all the wavelength-private encryption layers and shared decoding layers. Taking $L_2$ as an example, a slight rotation by merely 1° leads to a drastic PCC reduction from nearly 1 to below 0.05, with the plaintext fully invisible to the naked eye, as illustrated in Figure 3d. The PCC response curve for brute-force attacks via lateral displacement along the *x*- and *y*-directions, plotted in Figure 4a and Figure 4c for all the wavelength-private encryption layers and shared decoding layers, reveals similar results. A lateral displacement of $L_2$ by 10 μm (one pixel) along either the *x*- or *y*-direction reduces the PCC from nearly 1 to below 0.5, as illustrated in Figure 4b. These phenomena indicate that the plaintext information is thoroughly dissipated into chaotic background noise without retaining any recognizable plaintext contours, which fundamentally prevents adversaries from conducting reverse cracking.

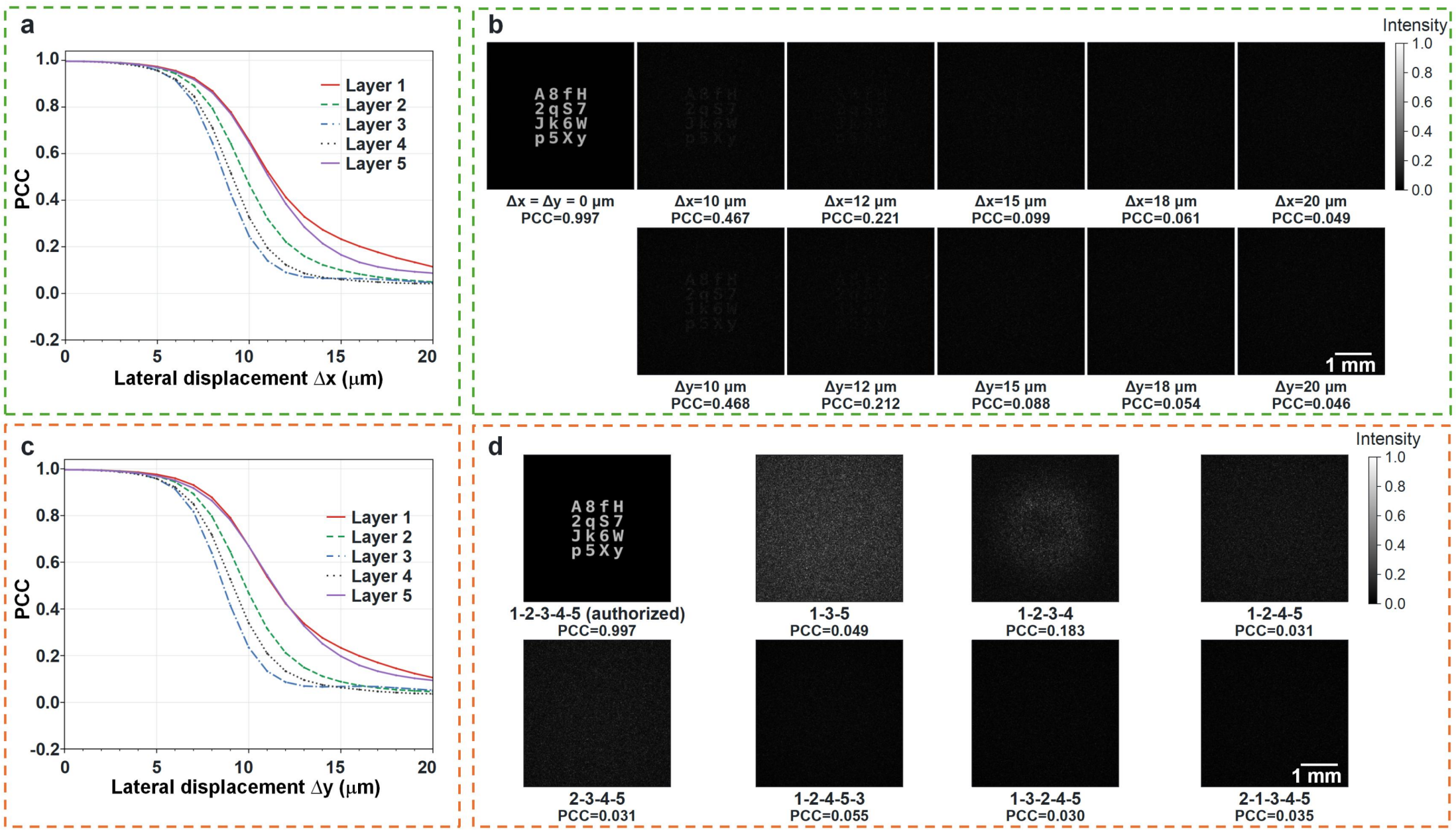

**Figure 4. Active contrast-dissipation under unauthorized physical keys: lateral misalignment and layer-stack tampering.** (a) Full-detector PCC versus lateral displacement Δ*x* of each of the five layers (0–20 μm). (b) Detector-plane intensity for Δ*x* = 10, 12, 15, 18 and 20 μm (PCC = 0.467, 0.221, 0.099, 0.061 and 0.049) and for the same displacements applied along the *y*-direction (PCC = 0.468, 0.212, 0.088, 0.054 and 0.046). A lateral displacement of one pixel (10 μm) reduces the PCC from nearly 1 to below 0.5, and a displacement of 15 μm to below 0.1. (c) Full-detector PCC versus lateral displacement Δ*y*. (d) Detector-plane intensity for the authorized stacking sequence $L_1$–$L_2$–$L_3$–$L_4$–$L_5$ (PCC = 0.997) and for seven structural tampering scenarios.

Finally, we conduct brute-force cracking tests on the stacking sequence of the wavelength-private encryption layers and shared decoding layers, the third core dimension of the physical keys, to further validate the efficacy of the proposed active contrast-dissipation mechanism. In this system, the standard authorized layer configuration is sequentially defined as $L_1$–$L_2$–$L_3$–$L_4$–$L_5$ as shown in Figure 4d. We systematically evaluate two typical structural tampering scenarios that are highly accessible for adversarial cracking: incomplete layer configurations and disordered layer stacking. Representative test cases include incomplete layer configurations of $L_1$-$L_3$-$L_5$ (missing $L_2$ and $L_4$), $L_1$-$L_2$-$L_3$-$L_4$ (missing $L_5$), $L_1$-$L_2$-$L_4$-$L_5$ (missing $L_3$), $L_2$-$L_3$-$L_4$-$L_5$ (missing $L_1$) as well as the disordered stacking sequence of $L_1$-$L_2$-$L_4$-$L_5$-$L_3$, $L_1$-$L_3$-$L_2$-$L_4$-$L_5$ and $L_2$-$L_1$-$L_3$-$L_4$-$L_5$ in Figure 4d. Experimental results show that the target plaintext completely disappears visually under structural mismatch conditions. This phenomenon is also attributed to the active contrast-dissipation effect, which thoroughly dissipates valid plaintext information into chaotic background noise without leaving any plaintext contours, thereby fundamentally preventing adversaries from performing reverse cracking. Quantitative results demonstrate that the native PCC drops sharply from 0.997 to nearly zero. The complete visual vanishing of plaintext and the near-zero PCC further verify the outstanding anti-cracking performance and robustness of the proposed active contrast-dissipation strategy against stacking sequence brute-force attacks.

## 3 Discussion & Conclusion

### 3.1 Fabrication Tolerance and Optical Nonlinearity Analysis

The proposed all-optical encryption system adopts a five-layer architecture consisting of three wavelength-private encryption layers and two shared decoding layers. Each diffractive layer comprises 400×400 units with a lateral pitch of 10 μm, covering an active aperture of 4 mm×4 mm. The continuous surface-height profiles of these layers are uniformly quantized over a height range of 0–7 μm. This geometry is well suited to fused-silica photolithography and multilevel reactive-ion etching [22, 23]. Therefore, we further analyze the influence of fabrication quantization, as shown in Figure 5. Increasing the quantization resolution to 32 (5-bit) levels improves the PCC of the reconstructed plaintext to 0.971. Further increasing the levels to 64 (6-bit), 128 (7-bit), and 256 (8-bit) yields PCC values of 0.993, 0.996, and 0.997, respectively. Although finer quantization delivers reconstruction performance approaching that of the continuous-height counterpart, it substantially increases fabrication difficulty and cost. By comprehensively balancing optical reconstruction fidelity, nanofabrication complexity, and manufacturing expense, we select 32 (5-bit) quantization levels as a practical resolution for prototype fabrication.

Introducing optical nonlinearity offers a promising route to further strengthen the physical-layer security of the cryptosystem. To mitigate risks from coherent complex-field interception, future incorporation of two-dimensional nonlinear materials (e.g., monolayer $MoS_2$ [24] or graphene) into the shared layers could introduce intensity-dependent optical responses while bridging diffractive cryptoprocessors with all-optical neural computing. In addition, integrating anisotropic subwavelength metasurfaces [26, 27] would enable polarization and spatial-mode multiplexing, expanding the physical access-control dimensions whose security enhancements can be systematically quantified against targeted attack models. Such multi-dimensional physical degrees of freedom will further enrich the multi-dimensional physical keys and reinforce the active contrast-dissipation mechanism for high-security all-optical encryption.

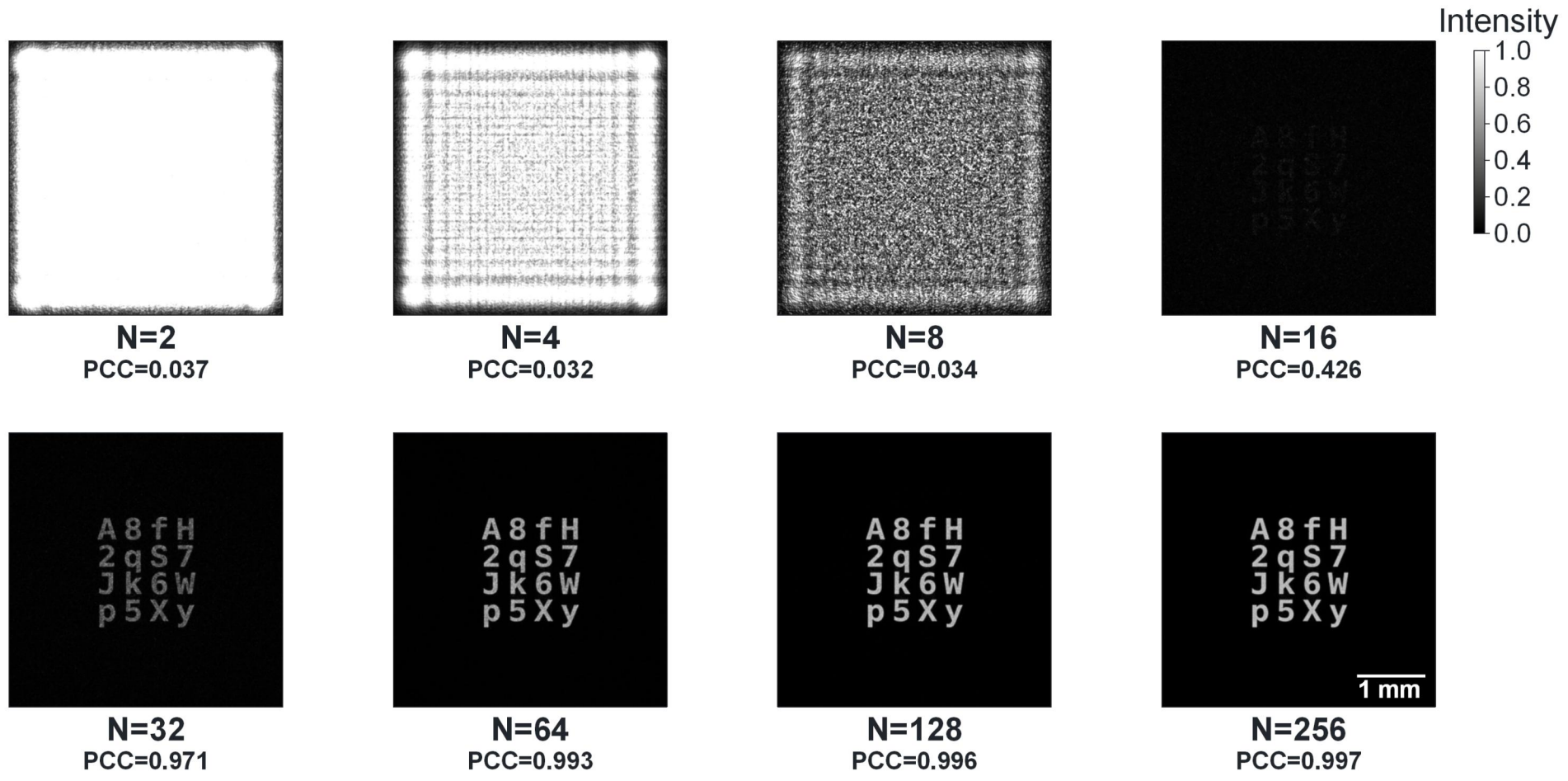

**Figure5. Fabrication tolerance: reconstruction fidelity versus quantization resolution.** Detector-plane intensity and PCC of the composite plaintext when the continuous 0–7 μm height profiles are uniformly quantized to $N = 2, 4, 8, 16, 32, 64, 128$ and 256 levels (1–8 bit).

### 3.2 Comparison with Conventional Diffractive Encryption System

We further compare our system with conventional diffractive optical encryption approaches [9]. Lacking the active contrast-dissipation mechanism, traditional architectures tend to expose recognizable plaintext or its contours under unauthorized physical keys, such as wavelength detuning (Figure 6a) and in-plane rotation (Figure 6b). For instance, clear plaintext contours can still be identified under a wavelength detuning of ±20 nm and an in-plane rotation of 5°. This contrast highlights the superiority of the proposed active contrast-dissipation strategy in this work.

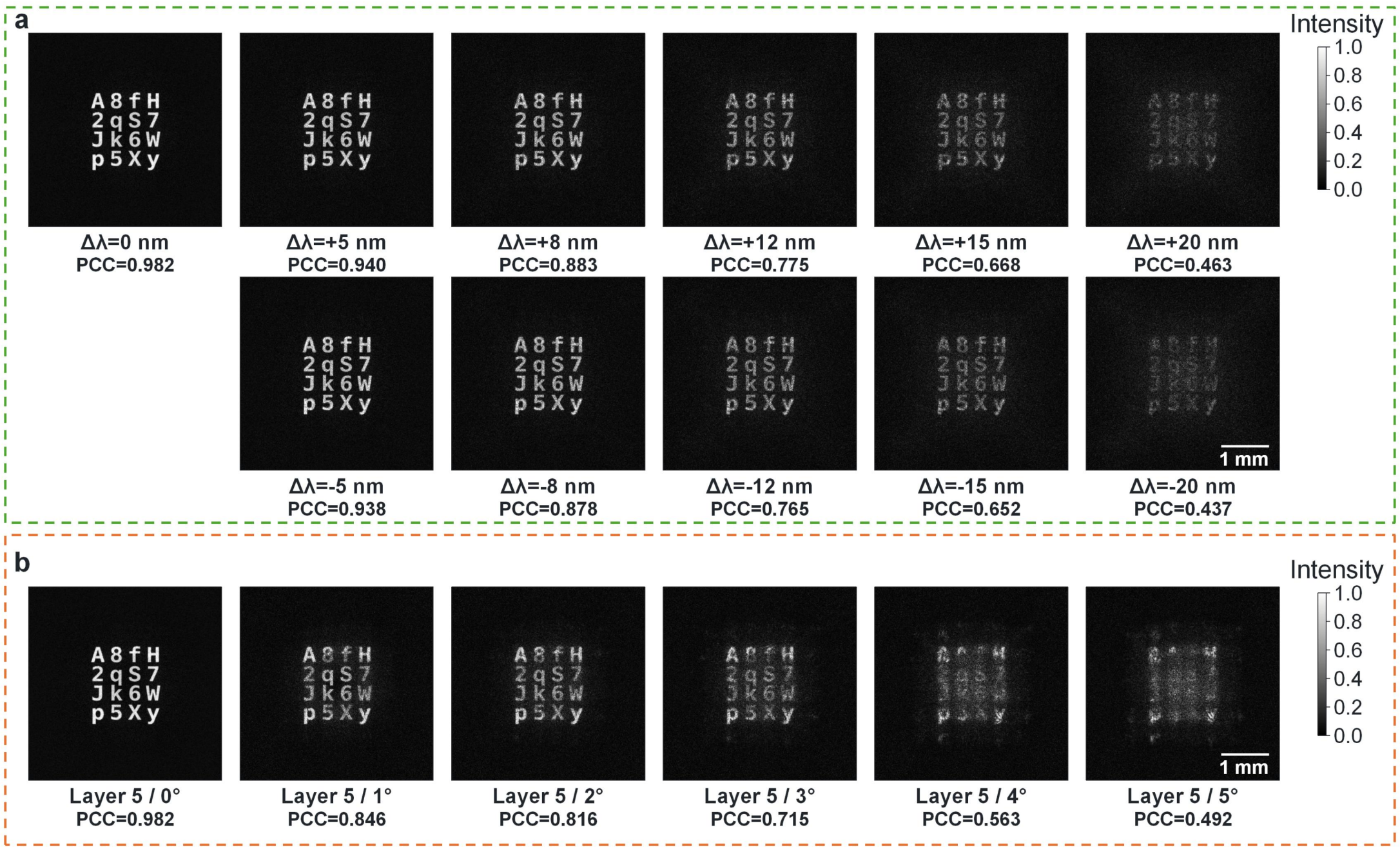

**Figure 6. Performance of conventional approaches under typical spoofing attacks.** (a) Detector-plane intensity for wavelength detunings Δλ. Clear plaintext contours can still be identified even at Δλ=±20 nm. (b) Detector-plane intensity for in-plane rotation. The plaintext contours remain identifiable even when the in-plane rotation is 5°.

### 3.3 Conclusion

In this work, we propose and validate an active contrast-dissipation all-optical encryption framework built upon a special designed "*M*+*N*" decoupled multi-wavelength diffractive deep neural network. By physically partitioning wavefront modulation into *M* wavelength-private encryption layers and *N* shared decoding layers, the architecture abandons the fully coupled design of conventional multi-wavelength encryptors and enables wavelength-private information isolation together with fine-grained hierarchical access control. For the implemented configuration of *M*=3 and *N*=2, the three ID patterns are reconstructed with full-detector PCCs of 0.976, 0.969 and 0.982 under their pre-assigned illumination wavelengths, whereas mismatched wavelengths leave the corresponding identities inactive (PCC ≈ 0). With all three wavelengths illuminating the system simultaneously and the multi-dimensional physical keys correctly configured, the target composite plaintext is recovered with a PCC of 0.997, confirming that high-fidelity authorized decryption is fully preserved.

Departing from conventional passive single-objective forward-fidelity optimization, the proposed active contrast-dissipation mechanism imposes an explicit constraint on outputs under physical perturbation: it actively suppresses the foreground-background intensity contrast, such that any deviation from authorized physical keys dissipates the plaintext into randomized, diffuse speckle rather than leaving behind recognizable residual features. Systematic brute-force tests covering the three dimensions of the multi-dimensional physical key, including the wavelength detuning, spatial misalignment, and disordered layer stacking, to verify this behavior, accompanied by a marked decline in PCC. Across all unauthorized test configurations, the plaintext vanishes, and no recoverable contour persists, suppressing residual semantic leakage which is a critical vulnerability that renders traditional passive diffractive encryptors prone to inversion attacks. Overall, the proposed active contrast-dissipation all-optical encryption achieves both exceptional anti-attack capability and practical engineering feasibility, offering a promising route toward advanced physically secure optical encryption.

## 4 Methods

### 4.1 Angular-Spectrum Propagation Model

To model optical propagation and its response to physical perturbations, inter-layer optical propagation is modeled using non-paraxial scalar angular-spectrum theory with finite-aperture boundary constraints. Each layer contains an active modulation area of 4 mm×4 mm (400×400 pixels, 10 $\mu$m pitch) enclosed by an opaque absorbing frame that blocks unmodulated direct transmission. The axial spacing between each layers is fixed at $z$=5 cm, while the final decoding layer ($L_5$) to the detector plane is set to 10 cm. Let $D$ = 4mm denote the full side length of the square active area and $a = D/2 = 2$ mm denote its half-width. Based on optical diffraction criteria, the critical Rayleigh distance $Z_f$ is defined as:

$$Z_f=\frac{2D^2}{\lambda} \tag{1}$$

For the central wavelength $\lambda_2$=532 $nm$, $Z_f \approx 60.1\ m$. Since the actual inter-layer spacing satisfies $z \ll Z_f$, the optical field operates within the near-field diffraction regime. At this scale, the Fresnel number $F$, which is the governing dimensionless parameter characterizing near-field diffraction can be calculated as:

$$F=\frac{a^2}{\lambda z}\approx 150 \gg 1 \tag{2}$$

The condition $F \gg 1$ places the modeled aperture in the near-field regime. To model wave propagation while accommodating large diffraction angles, inter-layer diffraction is described by non-paraxial scalar angular-spectrum theory. Let $U_l(x, y; \lambda)$ denote the complex optical field incident on the $l$-th layer. The plate modulates the wavefront via the modulation unit height $\mathrm{h}_l(x, y) \in [0, \mathrm{h}_{\max}]$ and $\mathrm{h}_{\max}$=7 μm, where the chromatic phase-delay profile $\phi_l(x, y; \lambda)$ is determined by the material dispersion of fused silica:

$$\phi_l(x, y; \lambda)=\frac{2\pi}{\lambda}[n_{\mathrm{silica}}(\lambda)-n_{\mathrm{air}}]\mathrm{h}_l(x, y) \tag{3}$$

Spatial truncation is enforced by the binary aperture function $A_l(x, y)$ corresponding to the opaque absorbing frame. The field immediately exiting the $l$-th diffractive layer is given by:

$$U_l^{+}(x, y; \lambda)=A_l(x, y) \exp\left[i\phi_l(x, y; \lambda)\right] U_l(x, y; \lambda) \tag{4}$$

The modulated field propagates across the axial distance $z$ to the next plane. The complex amplitude distribution $U_{l+1}(x, y; \lambda)$ is evaluated via spatial-frequency transfer:

$$U_{l+1}(x, y; \lambda)=\mathrm{F}^{-1}\left\{\mathrm{F}\left[U_l^{+}(x, y; \lambda)\right] \cdot H\left(f_x, f_y; z, \lambda\right)\right\} \tag{5}$$

where F and $\mathrm{F}^{-1}$ represent the two-dimensional discrete Fourier transform and its inverse, implemented with a fast Fourier transform (FFT), and $\left(f_x, f_y\right)$ are the transverse spatial-frequency coordinates. Defining the normalized transverse-frequency parameter $q=\left(\frac{\lambda}{n_{\mathrm{air}}}\right)^2\left(f_x^2+f_y^2\right)$, the exact angular-spectrum transfer function $H\left(f_x, f_y; z, \lambda\right)$ is formulated as:

$$H\left(f_x, f_y; z, \lambda\right)=\begin{cases}\exp\left[i\frac{2\pi\, n_{\mathrm{air}}\, z}{\lambda}\left(\sqrt{1-q}\right)\right], & q\leq 1\\ \exp\left[-\frac{2\pi\, n_{\mathrm{air}}\, |z|}{\lambda}\sqrt{q-1}\right], & q>1\end{cases} \tag{6}$$

Here, $q\leq 1$ corresponds to propagating components, whereas $q > 1$ corresponds to evanescent components that decay exponentially along the forward propagation direction $(z\geq 0)$. On the sampling grid used in this study, the maximum value of $q$ is below 1; therefore, only the propagating branch is involved in the numerical calculations. In our numerical implementation, the common carrier phase $\exp\left(i2\pi n_{\mathrm{air}}z/\lambda\right)$ is factored out from the transfer function. The phase of the resulting propagating kernel is evaluated using the algebraically equivalent expression $-\frac{2\pi n_{\mathrm{air}} z}{\lambda}\frac{q}{1+\sqrt{1-q}}$ to reduce subtractive-cancellation errors in single-precision floating-point arithmetic.

### 4.2 Active Contrast-Dissipation Mechanism

The active contrast-dissipation mechanism relies on a dual-path optimization framework consisting of an independent credential-verification mode and a cooperative composite-plaintext mode, which is formulated to coordinate multi-task hierarchical access control while suppressing the target contrast under unauthorized physical keys.

Under the credential-verification mode, each single wavelength $\lambda_i$ ( $i\in\{1, 2, 3\}$ ) exclusively traverses its dedicated private layer $L_i$ and the common decoding layers $(L_4, L_5)$. The dedicated identity credential loss $\mathrm{L}_{\mathrm{ID}}$ is governed by the following equation: the identity objective uses weighted mean-squared error (WMSE) and the structural similarity index (SSIM), both defined below.

$$L_{ID}=\sum_{i=1}^{M}\left[100\cdot\mathrm{WMSE}\left(I_{\lambda_i}, T_{ID_i}\right)+20\cdot\left(1\text{-}SSIM\left(I_{\lambda_i}, T_{ID_i}\right)\right)\right] \tag{7}$$

where $I_{\lambda_i}$ and $T_{\mathrm{ID}_i}$ are the normalized output intensity and the ground-truth identity pattern of the $i$-th ID.

For an output intensity image $I$ and its target image $T$, the weighted mean-squared error, denoted by WMSE, is defined as:

$$\mathrm{WMSE}(I,T)=\frac{\sum_{p=1}^{N_{\mathrm{pix}}} w_p\left(I_p-T_p\right)^2}{\sum_{p=1}^{N_{\mathrm{pix}}} w_p} \tag{8}$$

where $N_{\mathrm{pix}}$ is the total number of evaluated pixels, $I_p$ and $T_p$ denote the output intensity and target value at pixel $p$, respectively, and $w_p \geq 0$ is the corresponding pixel weight, with $\sum_{p=1}^{N_{\mathrm{pix}}} w_p > 0$. The weight map controls the contributions of different spatial regions to the reconstruction error. WMSE reduces to the conventional mean-squared error when all weights are equal.

The structural similarity index (SSIM) measures the similarity between the output and target images by comparing their local luminance, contrast, and structure. For a local window centered at pixel $p$, SSIM is defined as:

$$\mathrm{SSIM}_p(I,T)=\frac{\left(2\mu_{I,p}\mu_{T,p}+C_1\right)\left(2\sigma_{IT,p}+C_2\right)}{\left(\mu_{I,p}^2+\mu_{T,p}^2+C_1\right)\left(\sigma_{I,p}^2+\sigma_{T,p}^2+C_2\right)} \tag{9}$$

where $\mu_{I,p}$ and $\mu_{T,p}$ are the local means, $\sigma_{I,p}^2$ and $\sigma_{T,p}^2$ are the local variances, and $\sigma_{IT,p}$ is the local covariance. All local statistics are calculated using the same normalized window weights. The stabilization constants are defined as $C_1=(K_1L)^2$ and $C_2=(K_2L)^2$, where $K_1$ and $K_2$ are positive dimensionless constants and $L$ is the prescribed image intensity dynamic range. The overall SSIM is obtained by averaging the scores over all valid local windows:

$$\mathrm{SSIM}(I,T)=\frac{1}{N_{\mathrm{win}}}\sum_{p=1}^{N_{\mathrm{win}}}\mathrm{SSIM}_p(I,T) \tag{10}$$

where $N_{\mathrm{win}}$ is the number of valid windows. The corresponding structural similarity loss is defined as $1-\mathrm{SSIM}(I,T)$.

Under the cooperative composite-plaintext mode, multi-wavelength beams concurrently pass through the full cascade system. Because the multi-wavelength carriers are mutually temporally incoherent, the composite-plaintext intensity registered by the square-law detector is governed by an incoherent summation:

$$I_{comp}(x,y)=\sum_{i=1}^{M}|U_{out}(x,y;\lambda_i)|^2 \tag{11}$$

where $M$ denotes the number of wavelengths.

The intensity readout is nonlinear in the complex field. Therefore, the composite-plaintext baseline reconstruction loss $L_{\text{plain, base}}$ is formulated as:

$$L_{\text{plain,base}}=100\cdot\text{WMSE}(I_{\text{comp}},T_{\text{plain}})+100\cdot[1\text{-SSIM}(I_{\text{comp}},T_{\text{plain}})]+ \\ 50\cdot[1\text{-PCC}(I_{\text{comp}},T_{\text{plain}})]+L_{\text{peak}} \tag{12}$$

where WMSE, SSIM, and PCC denote the weighted mean-squared error, structural similarity index, and Pearson correlation coefficient computed directly on raw floating-point detector intensities without normalization.

To suppress the target contrast under unauthorized physical keys, an asymmetric peak-suppression penalty $L_{\text{peak}}$ replaces conventional polynomial binarization:

$$L_{\text{peak}}=\alpha_{\text{peak}}\frac{1}{N_{\text{pix}}}\sum_{x,y}\left[\text{ReLU}\left(I_{\text{comp}}(x,y)\text{-}I_{\text{th}}\right)\right]^2 \tag{13}$$

where ReLU denotes the rectified linear unit, $N_{\text{pix}}$ is the total pixel count, $\alpha_{\text{peak}}$=20, and the clipping threshold is set to $I_{\text{th}}$=1.1.

All simulations were performed on an NVIDIA GeForce RTX 4090 GPU (24 GB VRAM) using PyTorch 2.8.0, Python 3.12, CUDA 12.8, and Ubuntu 22.04. The complete simulation run required approximately 15 h, with a peak host-memory usage of approximately 13 GB.

### 4.3 Pearson Correlation Coefficient (PCC) Evaluation Protocol

The Pearson correlation coefficient (PCC) is used to quantify the global linear correlation between the reconstructed intensity distribution and the corresponding target pattern. PCC is computed directly from spatially matched floating-point intensity arrays, without smoothing or conversion to absolute correlation. Comparisons across different conditions use the same target pattern, evaluation region, and preprocessing procedure. Under this consistent evaluation protocol, a PCC closer to 1 indicates stronger positive linear correlation between the reconstructed and target intensity distributions. PCC is defined by:

$$\text{PCC}=\frac{\sum_{i=1}^{N}(x_i\text{-}\bar{x})(y_i\text{-}\bar{y})}{\sqrt{\sum_{i=1}^{N}(x_i\text{-}\bar{x})^2\sum_{i=1}^{N}(y_i\text{-}\bar{y})^2}} \tag{14}$$

Here, $x_i$ and $y_i$ denote the target and reconstructed intensities at the $i$-th corresponding pixel, respectively, and $N$ is the total number of pixels in the evaluation region. Their spatial means over that region are defined as $\bar{x}=N^{-1}\sum_{i=1}^{N}x_i$ and $\bar{y}=N^{-1}\sum_{i=1}^{N}y_i$. PCC is dimensionless and ranges from −1 to 1: a value of 1 indicates perfect positive linear correlation, 0 indicates no linear correlation, and −1 indicates perfect negative linear correlation. The sign of the calculated coefficient is retained. PCC is mathematically undefined when either intensity array has zero variance.

**Acknowledgments**

This work was supported by the National Natural Science Foundation of China (Grant No. 62505012) and the Research Action Plan 2026 of Beihang University (Grant No. KG21028601).

**Disclosures**

The authors declare that there are no financial interests, commercial affiliations, or other potential conflicts of interest that could have influenced the objectivity of this research or the writing of this paper.

**Code, Data, and Materials Availability**

The data, code, and simulation materials supporting the findings of this study are available from the corresponding authors upon reasonable request.

**References**

[1] Tataria H, Shafi M, Molisch A F, et al. 6G wireless systems: Vision, requirements, challenges, insights, and opportunities[J]. Proceedings of the IEEE, 2021, 109(7): 1166-1199.

[2] Atzori L, Iera A, Morabito G. The internet of things: A survey[J]. Computer netw orks, 2010, 54(15): 2787-2805.

[3] Sicari S, Rizzardi A, Grieco L A, et al. Security, privacy and trust in Internet of Things: The road ahead[J]. Computer networks, 2015, 76: 146-164.

[4] Daemen J, Rijmen V. The design of Rijndael: AES—the Advanced Encryption Sta ndard[M]. Berlin: Springer-Verlag, 2002. DOI: 10.1007/978-3-662-04722-4.

[5] Rivest R L, Shamir A, Adleman L. A method for obtaining digital signatures and public-key cryptosystems[J]. Communications of the ACM, 1978, 21(2): 120-126.

[6] Diffie W, Hellman M E. New directions in cryptography[M]//Democratizing crypto graphy: the work of Whitfield Diffie and Martin Hellman. 2022: 365-390.

[7] Koblitz N. Elliptic curve cryptosystems[J]. Mathematics of computation, 1987, 48(1 77): 203-209.

[8] Shor P W. Algorithms for quantum computation: discrete logarithms and factoring [C]//Proceedings 35th annual symposium on foundations of computer science. Ieee, 1994: 124-134.

[9] Lin X, Rivenson Y, Yardimci N T, et al. All-optical machine learning using diffra

ctive deep neural networks[J]. Science, 2018, 361(6406): 1004-1008.
[10] Shen Y, Harris N C, Skirlo S, et al. Deep learning with coherent nanophotonic circuits[J]. Nature photonics, 2017, 11(7): 441-446.
[11] Refregier P, Javidi B. Optical image encryption based on input plane and Fourier plane random encoding[J]. Optics letters, 1995, 20(7): 767-769.
[12] Park C, Jeon Y, Lee S, et al. Recomposable Layered Metasurfaces for Wavelength-Multiplexed Optical Encryption via Modular Diffractive Deep Neural Networks[J]. Advanced Functional Materials, 2026, 36(24): e23309.
[13] Wen D, Yue F, Li G, et al. Helicity multiplexed broadband metasurface holograms[J]. Nature communications, 2015, 6(1): 8241.
[14] Kamali S M, Arbabi E, Arbabi A, et al. Angle-multiplexed metasurfaces: encoding independent wavefronts in a single metasurface under different illumination angles[J]. Physical Review X, 2017, 7(4): 041056.
[15] Ren H, Briere G, Fang X, et al. Metasurface orbital angular momentum holography[J]. Nature communications, 2019, 10(1): 2986.
[16] Pappu R, Recht B, Taylor J, et al. Physical one-way functions[J]. Science, 2002, 297(5589): 2026-2030.
[17] Mengu D, Luo Y, Rivenson Y, et al. Analysis of diffractive optical neural networks and their integration with electronic neural networks[J]. IEEE Journal of Selected Topics in Quantum Electronics, 2019, 26(1): 1-14.
[18] Gao Y, Jiao S, Fang J, et al. Multiple-image encryption and hiding with an optical diffractive neural network[J]. Optics Communications, 2020, 463: 125476.
[19] Wu Y, Duan H, He Y, et al. Diffractive optical encryption systems based on multiple wavelengths and multiple distances[C]//Photonics. MDPI, 2024, 11(10): 922.
[20] Mengu D, Zhao Y, Yardimci N T, et al. Misalignment resilient diffractive optical networks[J]. Nanophotonics, 2020, 9(13): 4207-4219.
[21] Coherent Corp. OBIS LX/LS laser data sheet[EB/OL]. [2026-11-06]. https:obis-family-ds.pdf.
[22] Tudor R, Bulzan G A, Kusko M, et al. Multilevel spiral axicon for high-order Bessel–Gauss beams generation[J]. Nanomaterials, 2023, 13(3): 579.
[23] Wlodarczyk K L, Mendez E, Baker H J, et al. Laser smoothing of binary gratings and multilevel etched structures in fused silica[J]. Applied optics, 2010, 49(11): 1997-2005.
[24] Sahu M C, Sahoo S, Mallik S K, et al. Multifunctional 2D MoS2 optoelectronic artificial synapse with integrated arithmetic and reconfigurable logic operations for in-memory neuromorphic computing applications[J]. Advanced Materials Technologies, 2023, 8(2): 2201125.
[25] Abdollahramezani S, Hemmatyar O, Adibi A. Meta-optics for spatial optical analog computing[J]. Nanophotonics, 2020, 9(13): 4075-4095.
[26] Chen H, Feng J, Jiang M, et al. Diffractive deep neural networks at visible wavelengths[J]. Engineering, 2021, 7(10): 1483-1491.
[27] Chen H, Shen Y. Large-scale distributed diffractive-interference hybrid photonic chiplets[J]. Advanced Photonics, 2024, 6(4): 040502-040502.